\documentclass[11pt]{article}
\usepackage{graphicx,amssymb,amsmath}
\sloppypar
\def\etal{{et~al.}}
\def\Ne{\ensuremath{N_{\rm e}}}
\def\Nreal{\ensuremath{N_\mathrm{real}}}
\def\Nbg{\ensuremath{N_\mathrm{bg}}}
\def\Ncef{\ensuremath{N_\mathrm{CEF}}}
\def\Neas{\ensuremath{N_\mathrm{EAS}}}
\def\Pbc{\ensuremath{P_\mathrm{bc}}}
\def\PN{\ensuremath{P_N}}
\DeclareMathOperator{\median}{median}
\newcommand{\art}[5]{#1, #2 \textbf{#3}, #5 (#4).}
\newcommand{\artx}[6]{#1, #2 \textbf{#3}, #5 (#4); \texttt{arXiv:#6}.}
\newcommand{\prep}[3]{#1, \textit{#2}, \texttt{arXiv:#3}.}
\newcommand{\proc}[4]{#1, \textit{#2} (#3),        p.~#4.}
\newcommand{\book}[3]{#1, \textit{#2} (#3).}

\newcommand{\NEW}[1]{#1}
\begin{document}
\title{\bfseries
Region of Excessive Flux of PeV Cosmic Rays in the Direction
Toward Pulsars PSR J1840+5640 and LAT PSR J1836+5925
}

\author{%
G. V. Kulikov, M. Yu.\ Zotov\\[1mm]
D. V. Skobeltsyn Institute of Nuclear Physics,\\
M. V. Lomonosov Moscow State University,\\
Moscow 119992, Russia\\
\texttt{\{kulikov,zotov\}@eas.sinp.msu.ru}
}

\date{Version 2: April 22, 2010}
\maketitle
\begin{abstract}
An analysis of arrival directions of extensive air showers (EAS)
registered with the EAS MSU and EAS--1000 prototype arrays has revealed
a region of excessive flux of PeV cosmic rays in the direction
toward pulsars PSR J1840+5640 and LAT PSR J1836+5925 at significance
level up to~$4.5\sigma$.
The first of the pulsars was discovered almost 30 years ago and 
is a well-studied old radio-pulsar located at the distance of 1.7~pc
from the Solar system.
The second pulsar belongs to a new type of pulsars, discovered by the
space gamma-ray observatory Fermi, pulsations of which are not
observed in optical and radio wavelengths but only in the
gamma-ray range of energies (gamma-ray-only pulsars).
In our opinion, the existence of the region of excessive flux 
of cosmic rays registered with two different arrays provides a strong
evidence that isolated pulsars can give a noticeable contribution to the flux of
Galactic cosmic rays in the PeV energy range.
\end{abstract}


\section{Introduction}

The problem of the nature of the knee in the cosmic ray (CR) energy
spectrum at 3~PeV and, in a more generic aspect, of the origin of CRs in
the PeV energy range stays open for many years now. Attempts to collate
arrival directions of CRs with coordinates of their possible
astrophysical sources is one of the routes of investigations on the
subject. Another direction, which is being actively developed, is
observation of gamma rays from such sources.

In both approaches, the main subject of investigation are Galactic
supernova remnants (SNRs) since the model of diffusive shock acceleration
at the outer front of expanding SNRs has become widely spread, see, e.g.,
a review by Hillas 2005. This does not imply though that there are no
other astrophysical sources in the Galaxy that are able to accelerate
hadrons to PeV and higher energies. In particular, the first results
demonstrating that pulsars can act as effective CR accelerators appeared
soon after the discovery of pulsars and their identification with neutron
stars, see, e.g., Gunn~\& Ostriker 1969. After this, interest to pulsars
as possible sources of CRs with energies $\gtrsim10^{14}$~eV up to the
highest energies did not vanish, see, e.g., Blasi \etal\ 2000; Giller~\&
Lipski 2002; Bhadra 2003; Erlykin~\& Wolfendale 2004; Bhadra 2006. Models
considered included acceleration both in pulsar wind nebulae (e.g., the
Crab nebula) and by isolated pulsars. It was demonstrated in particular
that the Geminga pulsar is a possible candidate for being the single
source of the knee at 3~PeV (Bhadra 2003).

In our previous works, we have already reported on the number of regions
of excessive flux (REFs) found during an analysis of arrival directions
of EAS registered with the EAS MSU array and the EAS--1000 prototype
array (``PRO--1000'') and generated by CRs with energies of around PeV, 
(Kulikov~\& Zotov 2004; Zotov~\& Kulikov 2004, 2007, 2009). It was
demonstrated that there are no Galactic SNRs in the vicinity of some of
the REFs, though a considerable number of regions have isolated pulsars
nearby. In Zotov~\& Kulikov 2009, we pointed out that there are two
overlapping REFs in the vicinity of a point with equatorial coordinates
$\alpha=285^\circ$, $\delta=58^\circ$ in the EAS MSU and the PRO--1000
data sets. Taking into account possible errors in determination the
boundaries of the REFs, one could consider them as one and the same
region with the pulsar PSR J1840+5640 inside (coordinates of the pulsar
$\alpha\approx280.19^\circ$, $\delta\approx56.68^\circ$). There are
neither supernova remnants or SNR candidates inside or close to the
region, nor open clusters or OB-associations, which are sometimes
considered as possible sources of Galactic cosmic rays. PSR J1840+5640
was discovered back in 1980 (Shitov \etal\ 1980) and is a well-studied
old radio-pulsar of the age of $17.5\times10^6$ years located at 1.7~kpc
from the Solar system, see, e.g., Arzoumanian \etal\ 1994; Hobbs \etal\
2004.

This region attracted our attention once again after there appeared a
list of the brightest gamma-ray sources in the energy range from 100~MeV
to 100~GeV registered during the first three month of operation of the
Fermi Gamma-ray Space Telescope (Abdo \etal\ 2009a). It happened that
there is a pulsar (coordinates $\alpha\approx279.06^\circ$,
$\delta\approx59.46^\circ$) of a new type near the REF, a so called
gamma-ray-only pulsar, i.e., a pulsar that only ``blinks'' in gamma rays.
This pulsar as an intensive gamma-ray source was discovered by EGRET in
1991--1993 but the nature of the source remained unclear (Nolan \etal\
1994). It occurred to be the brightest unidentified gamma-ray source in
the energy ranges $>100$~MeV (significance level of $19.0\sigma$) and
$>1$~GeV ($14.2\sigma$) located far from the Galactic plane (Galactic
latitude $b\approx25^\circ$), see Hartman \etal\ 1999; Lamb \etal\ 1997.

Numerous attempts to elucidate the nature of the object named as 3EG
J1835+5918 and to find a definite astrophysical counterpart did not
succeed, see Mirabal \etal\ 2000; Reimer \etal\ 2001; Halpern \etal\
2002; Fegan \etal\ 2005; Halpern \etal\ 2007, and references therein.
After intensive studies of its error box in optical, X-ray, and radio
wave bands it was identified as an isolated neutron star correlating with
a faint X-ray source RX J1836.2+5925, discovered by ROSAT. It was
suggested that the object is a Geminga-like pulsar, which is not observed
in optical and radio wave bands, and an upper limit of the distance to
the object was obtained ($\lesssim800$~pc). All attempts to find
pulsations in the X-ray and gamma-ray energy ranges failed. Pulsations of
3EG J1835+5918 were not observed during the first year of observations by
the AGILE instrument either, though its gamma-ray emission was confirmed
at $15.6\sigma$ significance level (Bulgarelli \etal\ 2008; Pittori
\etal\ 2009). Only gamma-ray observations by the Fermi LAT lead to
unequivocal identification of 3EG J1835+5918 as a pulsar, named LAT PSR
J1836+5925.

Since the search for gamma-ray sources had become a natural part of the
search for PeV cosmic ray sources (Rowell \etal\ 2005), we performed a
more detailed analysis of the REF located in the direction to both
pulsars.

\section{Experimental Data}

Similar to our previous studies of EAS arrival directions, the main data
sets consist of 513~602 showers registered with the EAS MSU array in
1984--1990 and 1~342~340 showers of the PRO--1000 array registered in
1997--1999.  Technical description of the arrays were given by Vernov
\etal\ 1979, and Fomin \etal\ 1999, respectively. All EAS selected for
the analysis satisfy a number of quality criteria and have zenith angles
$\theta < 45.7^\circ$. Showers registered with the two arrays have
different number of charged particles~$\Ne$ in a typical event.  For the
EAS MSU array, the median value of~$\Ne$ is of the order of
$1.6\times10^5$, while that for the PRO--1000 array equals
$3.7\times10^4$. Assuming that the majority of the EAS was generated by
protons, one can estimate their energy as being at around 1~PeV.
\NEW{
Modern models of hadronic interactions and data on the chemical
composition of cosmic rays allow one to estimate energy of a primary
particle that caused an EAS.  The given values of~\Ne\ correspond to the
energies of a primary proton $E_0\approx1.7\times10^{15}$~eV and
$E_0\approx4.6\times10^{14}$~eV respectively, with an error of about
10\%--20\%.
}

The EAS MSU array allowed one to determine arrival directions of EAS with
a better accuracy than the PRO--1000 array but the mean error is
estimated to be of the order of~$3^\circ$ in both cases.

\section{Method of the Investigation}

The investigation is based on the method by Alexandreas \etal\
(Alexandreas \etal\ 1991), which has been used multiple times for the
analysis of arrival directions of CRs by different research groups. The
idea of the method is as follows. To every shower in the experimental
data set, arrival time of another shower is assigned in a pseudo-random
way. After this, new equatorial coordinates $(\alpha,~\delta)$ are
calculated for the ``mixed'' data set thus providing a ``mixed'' map of
arrival directions.  The mixed map differs from the original one but has
the same distribution in declination~$\delta$. In order to compare both
maps, one divides them into sufficiently small ``basic'' cells. A measure
of difference between any two regions (cells) of the two maps located
within the same boundaries is defined as \[ S = (\Nreal -
\Nbg)/\sqrt{\Nbg}, \] where \Nreal\ and \Nbg\ are the number of EAS
inside a cell in the real and ``mixed'' (background) maps
correspondingly. The mixing of the real map is performed multiple times,
and the mixed maps are averaged then in order to reduce the dependence of
the result on the choice of arrival times.  The method is based on an
assumption that the resulting mean ``background'' map has most of the
properties of an isotropic background, and presents the distribution of
arrival directions of cosmic rays that would be registered with the array
in case there is no anisotropy. Thus, deviations of the real map from the
background one may be assigned to a kind of anisotropy of arrival
directions of EAS registered at the array. As a rule, selection of
regions of excessive flux is performed basing on the condition $S>3$.

Similar to the paper Zotov~\& Kulikov 2009, basic cells are of the size
$0.5^\circ\times0.5^\circ$.  The number of cycles of mixing was increased
from 1000 to 10,000 in order to further improve the quality of the
background map.  Due to this, the difference between the basic cells of
any two consecutive averaged maps obtained close to the end of the mixing
reduced approximately by an order of magnitude and is
$\sim1$--2$\times10^{-3}$ for both data sets.

Cells of excessive flux (CEFs)\footnote{We use the word ``cell'' for
regions selected during the initial analysis and preserve the term
``region of excessive flux'' for regions formed of multiple cells.
Thus, contrary to cells, REFs are not necessarily quadrangles.}
of CRs were searched for in the following way.
Adjacent basic strips in~$\delta$ (each $0.5^\circ$ wide) were joined
into strips of width $\Delta\delta=3^\circ\dots30^\circ$ with step equal
to~$0.5^\circ$. Each wide strip was then divided into adjacent cells of
some fixed width~$\Delta\alpha$. After this, we calculated the number of
EAS inside each of these cells for both experimental (\Nreal) and
background (\Nbg) maps. For each pair of cells, $S$ was calculated. A
cell was considered as a possible CEF if $S>3$.
Every wide strip was divided into cells with a shift of $0.5^\circ$
in~$\alpha$ until all possible locations on the grid were covered.

Contrary to Zotov~\& Kulikov 2009, we did not restrict the search to
cells such that $\Delta\alpha=\Delta\delta/\cos\bar{\delta}$,
where~$\bar\delta$ is the mean value of~$\delta$ for the current strip,
and $\Delta\alpha$ is rounded to the nearest half-integer number, but
allowed $\Delta\alpha$ to deviate from that value by $\Delta\alpha/6$. As
a result, additional cells were considered.  For example, besides cells
of the size $\Delta\alpha\times\Delta\delta=3^\circ\times3^\circ$, we
also considered cells of the sizes $2.5^\circ\times3^\circ$ and
$3.5^\circ\times3^\circ$. Finally, only cells with more than 10 EAS
inside were analyzed.

The method of Alexandreas et~al. does not provide a direct answer to the
question about the chance probability of appearance of a CEF. It is
implicitly assumed that the deviation of~\Nreal\ from~\Nbg\ has a
Gaussian distribution, and thus one is expected to calculate the 
corresponding probability basing on the value of~$S$, which is an
estimate of the standard deviation of a sample and thus acts as a
significance level.  Hence, it is assumed the chance probability of a CEF
to appear is less than $1-0.9973$ providing it was selected at
significance level $S>3$.

In order to estimate the chance probability of appearance of a CEF basing
on the number of EAS inside, we introduced the following simple method
based on the binomial distribution (Zotov~\& Kulikov 2009). Let a shower
axis getting inside the CEF be a success. The number of trials equals the
number of showers~$N$ in the data set under consideration, and an
estimate of success (for a fixed region) equals $\tilde p = \Nbg/N$,
where~\Nbg\ is the number of showers in the cell of the background map.
The assumption is based on the fact that in the method of Alexandreas
\etal, \Nbg\ is considered as an expected number of showers in a cell.
Obviously, the chance probability of finding exactly~\Nreal\ EAS in a
cell (or region) equals
\[
	P(\nu=\Nreal) = C(N,\Nreal) \tilde{p}^{\Nreal} (1 - \tilde{p})^{N-\Nreal},
\]
where $\nu$ is a random variable equal to the number of successes in the
binomial model, and $C(N,\Nreal)$ is the corresponding binomial
coefficient. Obviously, it is more interesting to consider a probability
that there are at most~\Nreal\ showers in the cell $\PN=P(\nu\le\Nreal)$,
or its adjacent value $1-\PN$. The analysis performed and the data
presented below demonstrate that values of~\PN\ correlate well with the
values of chance probabilities calculated on the basis of significance
level~$S$.

The fact that $S>3$ for a given cell does not imply it is valid for its
sub-cells. More than this, since the method of Alexandreas \etal\ does
not anyhow restrict sizes and shapes of regions to be considered, the
non-uniformity of the distribution of data w.r.t.\ declination~$\delta$
can result in a selection of a CEF such that a great value of~$S$ is
obtained due to an excess of EAS in just one of its sub-cells.  A
situation such that $S<3$ for a number of adjacent cells but $S>3$ for a
cell (region) combined of them is possible. In order to improve the
robustness of selection of CEFs, we have tried a number of additional
quantities. One of the most efficient of them is the probability
$\Pbc=P(\xi<N_\mathrm{bc}^+)$, calculated from the following binomial
model. Let~$\xi$ be a random variable equal to the number of basic cells
of the given cell with an excess of EAS over the background values, and
$N_\mathrm{bc}^+$ is the corresponding experimental value. The number of
trials equals the number of basic cells in the CEF. It is natural to
assume the probability of success to be equal to~1/2. In the results
presented below, all CEFs satisfy a condition $\Pbc>0.9545$, which
corresponds to the significance level of $2\sigma$ for the Gaussian
distribution.  We thus reduced the chance that a CEF is selected solely
due to the non-uniformity of the EAS distribution w.r.t.~$\delta$.

\section{Results}

The region of excessive flux of CRs in the direction toward pulsars
PSR J1840+5640 and LAT PSR J1836+5925 found by the analysis of the
main EAS MSU data set is shown in the upper panel of Figure~1.
The REF consists of 99 CEFs with $S>3.0$ located within the boundaries
$\alpha=277.5^\circ\dots289^\circ$, $\delta=55.5^\circ\dots61^\circ$.
Some parameters of the CEFs the region consists of are presented in
Table~1.  $\Pbc=0.989308$ for a CEF with the highest value of~$S$.
For this CEF, an excess of the experimental flux over the background one
takes place in 53 of 84 basic cells that form the CEF.
The set of showers within the REF has a slightly greater median value
of~\Ne\ than the whole data set.  For example, $\median\Ne\approx1.7\times10^5$
for 1508 EAS with $\alpha=277.5^\circ\dots289^\circ$,
$\delta=56^\circ\dots60^\circ$, and $\median\Ne\approx1.8\times10^5$
for 856 EAS within the region covered by CEFs with $S>4.3$.

\begin{figure}[!ht]
\centerline{\includegraphics[width=0.7\textwidth]{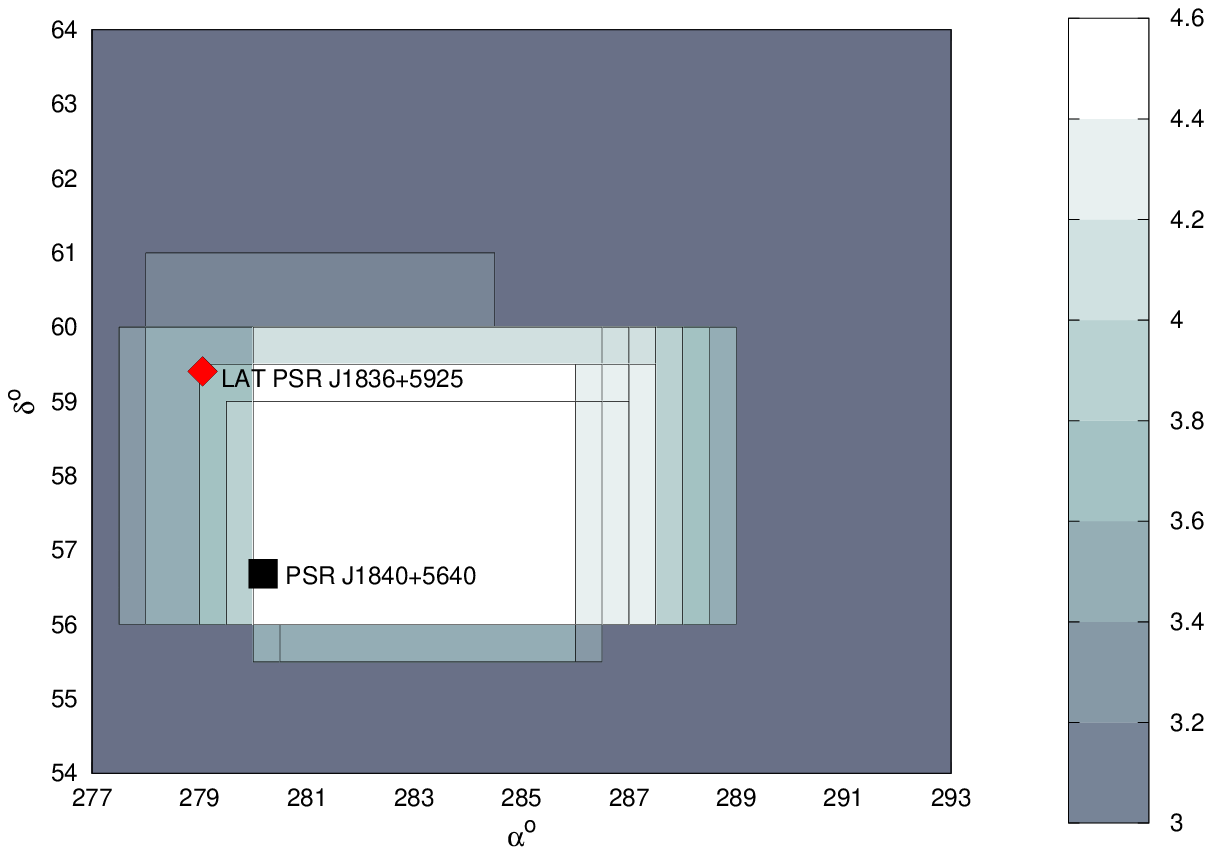}}
\centerline{\includegraphics[width=0.7\textwidth]{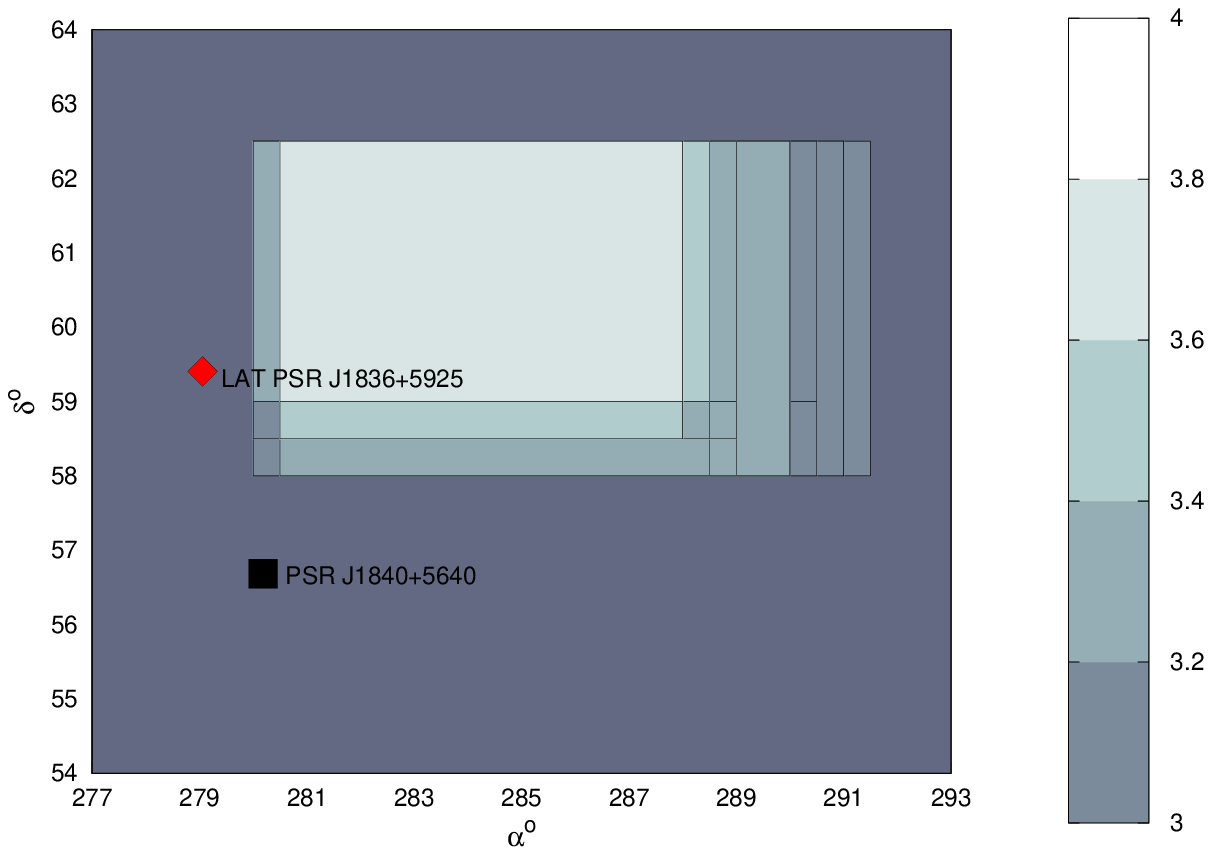}}
\caption{Upper panel: the region of excessive flux of CRs found
in the main data set of the EAS MSU array.  
Lower panel: the region of excessive flux of CRs found
in the PRO--1000 data set for showers with $\Ne>3\times10^4$.}
\end{figure}

\begin{table}[!ht]
\caption{%
Parameters of cells of excessive flux that form the REF found in the EAS
MSU data set (see Figure~1). Notation: $S$ is the significance level for
selecting CEFs, \Ncef\ is the number of CEFs selected for a given value
of~$S$, $\alpha$ and $\delta$ are the ranges of the corresponding
equatorial coordinates, \Neas\ is the range of the number of EAS
registered in selected CEFs, \Nbg\ is the corresponding range background
events, $\min\PN$ is the minimum value of~\PN\ for the CEFs selected.
}
\begin{center}
\begin{tabular}{|c|c|c|c|c|c|c|}
\hline
$S$ & \Ncef\ & $\alpha,\,^\circ$ & $\delta,\,^\circ$ & \Neas\ & \Nbg\ & $\min\PN$  \\
\hline
$>3.0$ & 99 & 277.5\dots289 & 55.5\dots61 & 518\dots1146 &449.6\dots1025.2& 0.998584 \\
$>3.5$ & 48 & 278\dots289   & 56\dots60   & 528\dots1146 &449.6\dots1025.1& 0.999735 \\
$>4.0$ & 21 & 280\dots287.5 & 56\dots60   & 537\dots1029 &451.5\dots904.3 & 0.999958 \\
$>4.3$ &  5 & 280\dots287   & 56\dots59.5 & 594\dots856  &496.6\dots736.8 & 0.999989 \\
4.478  &  1 & 280\dots286   & 56\dots59.5 & 744          &631.5         & 0.999994 \\
\hline
\end{tabular}
\end{center}
\end{table}

The result obtained for the whole EAS MSU data set is confirmed in the
case one selects subsets with an improved accuracy of determination of
arrival directions.  The most highly pronounced CEF for a set of showers
with $\Ne>10^5$ (343~141 EAS) has $S=3.951$, $\PN=0.999947$,
$\Pbc=0.993082$ and contains 620 EAS within the boundaries
$\alpha=280.5^\circ\dots287^\circ$, $\delta=54.5^\circ\dots58.5^\circ$
with the background level equal to 529.1 EAS. For this CEF, an excess of
the experimental flux over the background one takes place in 65 of 104
basic cells that form the CEF.

A similar result is obtained if one excludes showers that have arrival
directions close to the vertical.  In the set of showers with the zenith
angle $\theta>6^\circ$ (492~853 EAS), we found a CEF with $S=3.985$,
$\PN=0.999952$, $\Pbc=0.961815$ within the boundaries
$\alpha=280^\circ\dots286^\circ$, $\delta=56^\circ\dots59^\circ$. For the
CEF, an excess of the experimental flux over the background one takes
place in 44 of 72 basic cells. Thus, in all these cases we observe
regions of excessive flux with the probability to appear by chance that
can be estimated as $7.8\times10^{-5}$--$7.5\times10^{-6}$ if one uses
the maximum values of~$S$, or as $5.3\times10^{-5}$--$6\times10^{-6}$ in
case one uses the corresponding values of~\PN.

A region of excessive flux of CRs in the direction toward pulsars PSR
J1840+5640 and LAT PSR J1836+5925 was also found in the full data set of
the PRO--1000 array (with the significance level up to $S=3.573$ and
$\PN=0.999800$) and in its subsets similar to those described above.
Namely, we found a CEF with $S=3.785$, $\PN=0.999908$ in a subset with
$\theta>6^\circ$ (1~267~089 EAS), and a CEF with $S=3.603$ for a subset
with $\Ne>3\times10^4$ (471~554 EAS).  A REF found in the latter case is
shown in the lower panel of Figure~1. Some parameters of CEFs that form
the REF are presented in Table~2. The median value of~\Ne\ for the CEF
selected at the highest significance level equals $6.9\times10^4$. This
is slightly less than the corresponding value for the whole set of
showers with $\Ne>3\times10^4$ equal to $7.2\times10^4$.

\begin{table}[!ht]
\caption{%
Parameters of CEFs that form a region of excessive flux of CRs in the
subset of the PRO--1000 data with $\Ne>3\times10^4$. See notation in
Table~2.
}
\begin{center}
\begin{tabular}{|c|c|c|c|c|c|c|}
\hline
$S$ & \Ncef\ & $\alpha,\,^\circ$ & $\delta,\,^\circ$ & \Neas\ & \Nbg\ & $\min\PN$  \\
\hline
$>3.0$ & 57 & 280.0\dots291.5 & 58\dots62.5   & 544\dots1510  &476.9\dots1392.5  & 0.998526 \\
$>3.3$ & 10 & 280.5\dots288.5 & 58\dots62.5   & 695\dots1170  &612.3\dots1060.5  & 0.999472 \\
$>3.5$ &  3 & 280.5\dots288.5 & 59\dots62.5   & 810\dots917   &714.4\dots 816.4  & 0.999748 \\
3.603  &  1 & 280.5\dots288.0 & 59\dots62.5   & 865           &765.3             & 0.999811 \\
\hline
\end{tabular}
\end{center}
\end{table}

Thus, we have found cells of excessive flux in the experimental data of
the PRO--1000 array with the probabilities of appearing by chance that
can be estimated as 1.5--3.5$\times10^{-4}$ if we use the maximum values
of the significance level~$S$, or as 1--2$\times10^{-4}$ in case
estimates are based on the corresponding values of~\PN.

\NEW{
There is a possibility that coordinates of the two pulsars coincide
with the position of the REF just by chance.  We think a probability
of such a coincidence is small because the majority of known
pulsars are located in the vicinity of the Galactic plane while the
pulsars under consideration have galactic latitude $b\approx25^\circ$.
In particular, less than 20\% of more than 330 pulsars known in
the region $\delta>10^\circ$ have $b>20^\circ$, and only 26 of them
are located at distances less than 2~kpc from the Solar system
(Manchester et~al., 2005).
}

\NEW{
Flux of primary cosmic rays provides additional information about
the REF found.  One must take into account though that an area
covered by an EAS array does not coincide with the so called
``effective'' area.  The EAS MSU array covered an area of
approximately 0.5~km$^2$ but a circle of effective registration
was defined for the probability of registering a shower $\ge0.95$.
It has a radius equal to 20~m for showers with
$\Ne\ge2.5\times10^5$ ($E_0\gtrsim2.5$~PeV), 30~m for showers with
$\Ne\ge4\times10^5$ ($E_0\gtrsim3.7$~PeV), etc.
An area of effective registration for the PRO--1000 array was
defined less accurately thus we only provide data for the
EAS MSU array.
The flux of primary cosmic rays in the REF equals
$\approx4.4\times10^{-13}$~cm$^{-2}$~s$^{-1}$ for
showers with \Ne=2.5--4$\times10^5$ (237 EAS),
and $\approx2.3\times10^{-13}$~cm$^{-2}$~s$^{-1}$
for showers with \Ne=4$\times10^5$--10$^6$ (273 EAS.)
The number of showers in the REF with greater values of~\Ne\ 
is less than 100, thus we do not take it as statistically
sufficient.
}

\section{Discussion}

In spite of the fact that the boundaries of the regions of excessive flux
found in the data sets obtained with two different experimental arrays
are shifted with respect to each other, we think they point to one and
the same region.  Making this conclusion, we take into account the
discreetness of the grid and possible errors in determination of arrival
directions of EAS.

As we have already mentioned in the introduction, there are no other
Galactic objects usually considered as possible sources of PeV cosmic
rays in the vicinity of the REF described above. The list of objects
checked includes supernova remnants (Green 2009), including pulsar wind
nebulae (Roberts 2005), open clusters (Dias \etal\ 2002),
OB-associations, and SNR candidates (SIMBAD database).  Hence, a natural
question is whether pulsars PSR J1840+5640 and LAT PSR J1836+5925 are
able to accelerate protons and/or heavier nuclei up to energies of the
order of PeV.

To estimate the maximum energy of a particle accelerated in the wind near
the light cylinder of a pulsar, one can use the following expression
(Blasi \etal\ 2000): $E_{\max} = 0.34 \, Z \, B \, \Omega^2$~eV,
where $Z$ is the charge number of the particle, $B$ is the strength of
the surface magnetic field,~G, $\Omega$ is the angular velocity of a
pulsar, rad~s$^{-1}$, and the radius of the pulsar equals $10^6$~cm. The
typical energy of the accelerated particle equals approximately a half
of~$E_{\max}$.

For PSR J1840+5640, $B\approx1.59\times10^{12}$~G,
$\Omega\approx3.80$~rad~s$^{-1}$ (barycentric period $P\approx1.653$~s,
see Manchester \etal\ 2005). Thus, the maximum energy of a proton
accelerated by this pulsar equals approximately $7.8\times10^{12}$~eV. An
iron nuclei can be accelerated up to $2\times10^{14}$~eV.  It follows
that PSR J1840+5640 can contribute to the formation of the REF only in
case there is a considerable fraction of heavy nuclei in the flux of
primary CRs from the corresponding direction.

For LAT PSR J1836+5925, $B\approx0.5\times10^{12}$~G,
$\Omega\approx36.26$~rad~s$^{-1}$ (Abdo \etal\ 2009b). Thus, the maximum
energy of a proton accelerated by this pulsar is of the order of
$2.2\times10^{14}$~eV.  For an iron nuclei,
$E_{\max}\approx5.8\times10^{15}$~eV.  Hence, LAT PSR J1836+5925 is a
possible candidate for being an accelerator of charged particles in the
PeV energy range.

Let us assume basing on the estimates of~$E_{\max}$ that the REF
described above was formed due to the contribution of one or both
pulsars. Then there arises a question of how was the arrival direction of
the corresponding EAS kept so close to the direction to these pulsars. 
It is difficult to give a definite answer. On the one hand, we cannot
totally exclude a possibility that a part of showers that formed the REF
was initiated by photons but not hadrons. The PRO--1000 array did not
allow one to distinguish between gamma-ray and hadronic showers but the
EAS MSU array provided such an opportunity because it had a muon
detector. Since photonic showers contain very few muons in comparison
with hadronic ones, their fraction in the whole data set could be
estimated.  It was shown by Khristiansen \etal\ (1975) that the fraction of
photon-initiated showers in the data set obtained with the EAS MSU array
in 1960s was of the order of $10^{-3}$.
\NEW{%
A similar estimate was obtained basing on the results of the CASA-MIA
experiment, see Chantell \etal\ 1997.
Therefore, we have no ground to assume that the observed excess of
CRs is formed due to photons.
}

Another possibility is a negligible influence  of the Galactic magnetic
field (GMF) on particle trajectories, or even magnetic lensing.  It is
difficult to say anything definite about the transverse component of the
GMF along the line of sight to the pulsars but provide the usual
estimates of the mean values of the field in the vicinity of the Solar
system.  It is remarkable though that there exists a local magnetic field
going toward the Galactic longitude $l\approx90^\circ$, close to that of
both pulsars, with a strength about 2--3~$\mu$G (Manchester 1974). Known
values of the rotation measure RM and the dispersion measure DM of the
pulsar PSR J1840+5640 allow one to estimate the average value of the GMF
along the line of site to the pulsar (see, e.g., Han \etal\ 2006):
$\langle B_\parallel \rangle = 1.232 \, \mathrm{RM}/\mathrm{DM}$~$\mu$G.
For PSR J1840+5640, $\mathrm{RM} = -3$~rad~m$^{-2}$,
$\mathrm{DM}=26.698$~cm$^{-3}$~pc (Manchester \etal\ 2005). Thus,
$\langle B_\parallel \rangle\approx-0.138$~$\mu$G.

\NEW{
One might ask, why has not been the REF found in experimental data
of other arrays that studied PeV cosmic rays.
Actually, one can find a number of reports about small-scale
anisotropy (regions of excessive flux of CRs) in this energy range,
see, e.g., Benk\'o \etal\ 2004, and Sun \& Sun 1997.  Some of the
regions presented in these works have coordinates close to the other
regions found by us, but mostly the results do not agree.

On the other hand, a search for point-like sources of PeV CRs with
the data of CASA-MIA and KASCADE arrays gave a negative result, see
McKay \etal\ 1993, Antoni \etal\ 2004.
A method close to that of Alexandreas \etal\ was used for the analysis
of the CASA-MIA data.  A comparison of the experimental and background
fluxes was performed for regions of the size
$0.8^\circ\times0.8^\circ/\cos\delta$.
An existence of at least four adjacent regions with an excess of
the number of registered showers over the background at $>2.8\sigma$
was demanded.
The KASCADE data set was studied with the original method by
Alexandreas \etal\ 
Only regions of the size $0.5^\circ\times0.5^\circ/\cos\delta$
were analysed.
It is possible that in both cases the choice of such small regions
was the main reason of the negative result:
the fraction of EAS generated by PeV photons is not sufficient
for providing a considerable excess of the observed flux over the
background,
and it is unlikely that the arrival direction
of a charged PeV particle is preserved with such a high precision.
In our opinion, it would be very interesting to revisit
the data of the CASA-MIA and KASCADE experiments and to study
regions of the size $\ge3^\circ\times3^\circ/\cos\delta$.
}

To conclude, we cannot unequivocally claim that the region of excessive
flux of PeV cosmic rays described above was formed due to the
contribution of any or both of the pulsars PSR J1840+5640 and LAT PSR
J1836+5925. Still, we think the existence of the REF in the direction
toward these pulsars is worth paying attention as a possible sign that
isolated pulsars provide a more noticeable contribution to the flux of
Galactic cosmic rays than it is sometimes assumed. This point of view is
supported by the fact that there are four more REFs in the EAS MSU data
set selected with significance level $S>4$ that can be associated with
pulsars which do not have known SNRs nearby. These REFs will be the
subject of a separate article.

\bigskip 

We thank Michael Dormody of the Fermi LAT team for kindly providing us
the paper by Abdo \etal\ 2009b,
\NEW{%
and Richard Manchester (ATNF) for clarifying a number of questions
about the Galactic magnetic field.
We also thank N.~N.~Kalmykov, V.~P.~Sulakov, and A.~V.~Shirokov
for useful discussions of various aspects of work of the EAS MSU
and PRO--1000 arrays.
}

Only free, open source software was used
for the investigation. In particular, all calculations were performed
with GNU Octave (Eaton \etal\ 2008) running in Linux. This research has
made use of the SIMBAD database, operated at CDS, Strasbourg, France. The
research was partially supported by the Russian Foundation for
Fundamental Research grant No.~08-02-00540


\section*{References}
\frenchspacing
\begin{enumerate}
\item
	\artx{Abdo A.A., Ackermann M., Ajello M.
	\etal}{Astrophys. J. Suppl.}{183}{2009}{46}{0902.1340}
	(2009a)

\item
	Abdo A.A., Ackermann M., Ajello M.
	\etal), Science Express, 2 July 2009 (2009b):
	\texttt{http://www.sciencemag.org/cgi/content/abstract/1175558}

\item
	\art{Alexandreas D.E., Berley D., Biller S. \etal}{Astrophys.~J.}%
	{1991}{383}{L53}

\item
   \artx{Antoni T., Apel W.D.,  Badea A.F.
   \etal)}{Astrophys.~J.}{608}{2004}{865}{astro-ph/0402656}

\item
	\art{Arzoumanian Z., Nice D.J., Taylor J.H. \etal}{Astrophys. J.}{422}{1994}{671}

\item
   \artx{Benk\'o G., Erd\"os G.,  Nikolsky S.I. \etal}{Izv. RAN,
   Ser. Fiz.}{69}{2004}{1599}{astro-ph/0502065}

\item
   \proc{Bhadra A.}{Proc. 28th ICRC, Tsukuba, Japan}{2003}{303}

\item
	\artx{Bhadra A.}{Astropart. Phys.}{25}{2006}{226}{astro-ph/0602301}

\item
	\artx{Blasi P., Epstein R.I., Olinto A.V.}{Astrophys. J.}%
	{533}{2000}{L123}{astro-ph/9912240}

\item
	\artx{Bulgarelli A., Tavani M., Caraveo P. \etal}%
	{Astron. Astrophys.}{489}{2008}{L17}{0808.3464}

\item
    \artx{Chantell M.C., Covault C.E., Cronin J.W. \etal)}{Phys.
    Rev. Lett.}{79}{1997}{1805}{astro-ph/9705246}

\item
	\art{Dias W.S., Alessi B.S., Moitinho A. \etal}{Astron.
	Astrophys.}{389}{2002}{871}
	\texttt{http://www.astro.iag.usp.br/\~{}wilton/clusters.txt}

\item
   \book{Eaton J.W., Bateman D., Hauberg S.}{GNU Octave Manual Version~3}%
	{Network Theory Ltd., United Kingdom, 2008}
	(\texttt{http://www.octave.org/})

\item
   \prep{Erlykin A.D., Wolfendale A.W.}{Are there pulsars in the
	knee?}{astro-ph/0408225}

\item
	\art{Fegan S.J., Badran H.M., Bond I.H. \etal}%
	{Astrophys. J.}{624}{2005}{638}

\item
   \proc{Fomin Yu.A., Igoshin A.V., Kalmykov N.N. \etal}{Proc. 26th
   ICRC, Salt Lake City, 1}{Ed. D. Kieda, M. Salamon, and B. Dingus, 1999}{286}

\item
   \art{Giller M., Lipski M.}{J. Phys. G}{28}{2002}{1275}

\item
   \artx{Green D.A.}{Bull. of the Astron. Soc. of
	India}{37}{2009}{in press}{0905.3699}
	\texttt{http://www.mrao.cam.ac.uk/surveys/snrs/}

\item
	\art{Gunn J.E., Ostriker J.P.}{Phys. Rev. Lett.}{22}{1969}{728}

\item
	\artx{Halpern J.P., Gotthelf E.V., Mirabal N. \etal}%
	{Astrophys. J.}{573}{2002}{L41}{astro-ph/0205442}

\item
	\artx{Halpern J.P., Camilo F., Gotthelf E.V.}%
	{Astrophys. J.}{668}{2007}{1154}{0707.1547}

\item
	\art{Han J.L., Manchester R.N., Lyne A.G. \etal}%
	{Astrophys. J.}{642}{2006}{868}

\item
	\art{Hartman R.C., Bertsch D.L., Bloom S.D. et al.}{Astrophys.~J. Suppl.}%
	{1999}{123}{79}

\item
	\art{Hillas A.M.}{J. Phys. G}{31}{2005}{R95}

\item
	\art{Hobbs G., Lyne A.G., Kramer M. \etal}{Mon. Not. Roy. Astron. Soc.}%
	{353}{2004}{1311}

\item
	\book{Khirstiansen G.B., Kulikov G.V., Fomin Yu.A.}{Cosmic Radiation
	of Very High Energies}{Moscow, Atomizdat, 1975}

\item
   \prep{Kulikov G.V., Zotov M.Yu.}{A search for outstanding sources
	of PeV cosmic rays: Cassiopeia A, the Crab Nebula, the Monogem Ring--But
	how about M33 and the Virgo cluster?}{astro-ph/0407138}

\item
	\art{Lamb R.C., Macomb D.J.}{Astrophys. J.}{488}{1997}{872}

\item
	\art{Manchester R.N.}{Astrophys. J.}{188}{1974}{637}

\item
   \art{Manchester R.N., Hobbs G.B., Teoh A. \etal}{Astrophys.~J.}%
   {2005}{129}{1993}
   \mbox{\texttt{http://www.atnf.csiro.au/research/pulsar/psrcat/}}

\item
   \art{McKay T.A., Borione A., Catanese M.
   \etal)}{Astrophys.~J.}{417}{1993}{742}

\item
	\artx{Mirabal N., Halpern J.P., Eracleous  M. \etal}%
	{Astrophys. J.}{541}{2000}{180}{astro-ph/0005256}

\item
	\proc{Nolan P.L., Fierro J.M., Lin Y.C. et al.}{AIP Conf. Proc.,
	No. 304}{1994}{360}

\item
	\prep{Pittori C., Verrecchia F., Chen  A.W. \etal}{First AGILE
	Catalog of High Confidence Gamma-Ray Sources}{0902.2959v1}

\item
	\artx{Reimer O., Brazier K. T. S., Carrami\~nana A. \etal}%
	{Mon. Not. Roy. Astron. Soc.}{324}{2001}{772}{astro-ph/0102150}

\item
   \book{Roberts M.S.E.}{The Pulsar Wind Nebula Catalog (March 2005
	version)}{McGill University, Montreal, Quebec, Canada, available on the
	World-Wide-Web at \texttt{http://www.physics.mcgill.ca/\~{}pulsar/pwncat.html}}

\item
	\prep{Rowell G., Aharonian F., Plyasheshnikov A.}{Ground-Based
	Gamma-Ray Astronomy at Energies Above 10 TeV: Searching for
	Galactic PeV Cosmic-Ray Accelerators}{astro-ph/0512523}

\item
	SIMBAD database \texttt{http://simbad.u-strasbg.fr/simbad/}

\item
	\art{Shitov Y.P., Kuzmin A.D., Kutuzov S.M. \etal}%
	{Sov. Astron. Lett.}{6}{1980}{85}

\item
   \proc{Sun L., Sun S.}{Proc. 25th ICRC,
   Durban, South Africa, 4}{Ed. M.S.~Potgieter, B.C.~Raubenheimer, D.J.~van
   der~Walt, Potchefstroom, 1997}{165}

\item
	\proc{Vernov S.N., Khristiansen G.B., Atrashkevich~V.B. et al.}{Proc. 16th
	ICRC, Kyoto, 8}{1979}{129}

\item
   \art{Zotov M.Yu., Kulikov, G.V.}{Izvestiya RAN, ser.\ fiz.}{68}{2004}{1602}

\item
   \artx{Zotov M.Yu., Kulikov, G.V.}{Bull.\ Russ.\ Acad.\ Sci.: Physics}{71}{2007}{483}{astro-ph/0610944}


\item
   \artx{Zotov M.Yu., Kulikov G.V.}{Izvestiya RAN, ser.\ fiz.}{73}{2009}{612}{0902.1637}

\end{enumerate}
\end{document}